\documentclass[sigconf,balance=true, authorversion]{acmart}
\AtBeginDocument{%
  }

\usepackage{placeins}
\usepackage{hyperref}
\usepackage{url}
\usepackage[ruled,vlined,linesnumbered]{algorithm2e}
\usepackage{subcaption}
\usepackage{float}
\usepackage{xcolor}
\usepackage{soul}  
\usepackage{makecell}
\usepackage{multirow}
\definecolor{lightorange}{RGB}{255, 200, 100}
\definecolor{lightblue}{RGB}{173, 216, 230}
\definecolor{lightgreen}{RGB}{193, 255, 193}

\ifdefined\shownotes

  \newcommand{\customnote}[1]{%
    \sethlcolor{lightorange}\hl{#1}%
  }
  \newcommand{\yuxuannote}[1]{%
    \sethlcolor{lightblue}\hl{#1}%
  }
  \newcommand{\suggest}[1]{%
    \sethlcolor{yellow}\hl{#1}%
  }
  \newcommand{\trimbin}[1]{%
    \sethlcolor{red}\hl{#1}%
  }

  \newcommand{\jlrannote}[1]{%
    \sethlcolor{pink}\hl{#1}%
  }

\newcommand{\nt}[1]{%
    \sethlcolor{lightgreen}\hl{#1}%
  }
  
\else
  \newcommand{\customnote}[1]{}
  \newcommand{\yuxuannote}[1]{}
  \newcommand{\jlrannote}[1]{}
  \newcommand{\suggest}[1]{}
  \newcommand{\trimbin}[1]{}
  \newcommand{\nt}[1]{}

\fi
\newcommand{\tllpaper}{\emph{Lossless Pruning }}
\newcommand{\colbert}{ColBERT }
\newcommand{\colbertvtwo}{ColBERTv2}
\newcommand{\colbertee}{ColBERTv2-e2e}
\newcommand{\colbertrr}{ColBERTv2-rerank}

\newcolumntype{L}[1]{>{\raggedright\arraybackslash}p{#1}}
\newcolumntype{C}[1]{>{\centering\arraybackslash}p{#1}}
\newcolumntype{R}[1]{>{\raggedleft\arraybackslash}p{#1}}

\copyrightyear{2026}
\acmYear{2026}
\setcopyright{cc}
\setcctype{by}
\acmConference[SIGIR '26]{Proceedings of the 49th International ACM SIGIR Conference on Research and Development in Information Retrieval}{July 20--24, 2026}{Melbourne, VIC, Australia}
\acmBooktitle{Proceedings of the 49th International ACM SIGIR Conference on Research and Development in Information Retrieval (SIGIR '26), July 20--24, 2026, Melbourne, VIC, Australia}
\acmDOI{10.1145/3805712.3809726}
\acmISBN{979-8-4007-2599-9/2026/07}

\begin{document}

\title{A Voronoi Cell Formulation for Principled Token Pruning in Late-Interaction Retrieval Models}

\author{Yash Kankanampati}
\affiliation{%
  \institution{Université Sorbonne Paris Nord, CNRS, LIPN}
  \city{Villetaneuse}
  \country{France}
}
\email{kankanampati[at]lipn.fr}

\author{Yuxuan Zong}
\affiliation{%
  \institution{Sorbonne Université, CNRS, ISIR}
  \city{Paris}
  \country{France}
}
\email{yuxuan.zong[at]isir.upmc.fr}

\author{Nadi Tomeh}
\affiliation{%
  \institution{Université Sorbonne Paris Nord, CNRS, LIPN}
  \city{Villetaneuse}
  \country{France}
}
\email{tomeh[at]lipn.fr}

\author{Benjamin Piwowarski}
\affiliation{%
  \institution{Sorbonne Université, CNRS, ISIR}
  \city{Paris}
  \country{France}
}
\email{benjamin.piwowarski[at]cnrs.fr}

\author{Joseph Le Roux}
\affiliation{%
  \institution{Université Sorbonne Paris Nord, CNRS, LIPN}
  \city{Villetaneuse}
  \country{France}
}
\email{leroux[at]lipn.fr}
\renewcommand{\shortauthors}{Kankanampati et al.}

\begin{abstract}
Late-interaction models such as ColBERT offer competitive performance across various retrieval tasks but require storing a dense embedding for each document token, leading to a substantial index storage overhead. Past works address this by attempting to prune low-importance token embeddings based on statistical and empirical measures, but they often either lack formal grounding or are ineffective. To address these shortcomings, we introduce a framework grounded in hyperspace geometry and cast token pruning as a Voronoi cell estimation problem in the embedding space. By interpreting each token's influence as a measure of its Voronoi region, our approach enables principled pruning that retains retrieval quality while reducing index size. Through our experiments, we demonstrate that this approach serves not only as a competitive pruning strategy but also as a valuable tool for improving and interpreting token-level behavior within dense retrieval systems.



\end{abstract}



\maketitle

\section{Introduction}

\suggest{I propose to do a final overall check of our notation to be consistent. 1) I have seen several appearance of MS~MARCO and MS~MARCO. I propose to use MS~MARCO everywhere as in the original paper it has the space.  2) $\mathrm{Error}(...)$ or $Error(...)$}

\yuxuannote{Maybe need to mentioned that the voronoi pruning allow the model to be evaluate purely on the normalized vectors, while sigir2025 paper's vectors are bounded by norm 1 instead of normalized. Thus, our methods can be retrieved efficiently using PLAID or other search engines, while it could be hard for sigir2025 work (Say why). (Currently i don't know where to put actually. )}
\customnote{Y: This could maybe go in the conclusion?}
\customnote{B: yes}

Information retrieval has seen significant advances with the emergence of late-interaction models such as  \colbert{}~\cite{khattabColBERTEfficientEffective2020a, santhanamColBERTv2EffectiveEfficient2022a} and COIL~\cite{gaoCOILRevisitExact2021}, which represent documents and queries as sets of token-level embeddings and compute relevance via fine-grained token interactions, capturing lexical and semantic matches that are often missed by single-vector encoders such as DPR~\cite{karpukhin2020dense}. This expressivity enables strong retrieval effectiveness while maintaining competitive latency~\cite{santhanamPLAIDEfficientEngine2022}.

However, this performance comes at a substantial cost. A late-interaction model stores an embedding for every document token, resulting in indexes that are orders of magnitude larger than those of sparse or single-vector dense retrievers. In practice, this storage overhead becomes a fundamental bottleneck for large-scale retrieval, limiting deployability and increasing both memory and compute costs.

To reduce the index size, several works propose index compression techniques. For instance, \colbertvtwo~\cite{santhanamColBERTv2EffectiveEfficient2022a} leverages a linear projection to reduce the vector representation to a smaller dimension. Meanwhile, it also leverages the residual quantization to further reduce the storage required for each document token vector. 

However, not all token embeddings contribute equally to retrieval. Many tokens are weakly informative, only relevant for vanishingly small sets of queries or encode redundant information~\citep{godeyAnisotropyInherentSelfAttention2024}, suggesting that pruning uninformative tokens can reduce storage while minimally impacting retrieval performance. Conceptually, this is analogous to how removing the token “The” from the sentence “The capital of France is Paris” would have virtually no impact on the information conveyed. 

Existing approaches attempt to prune tokens by typically relying on heuristics or learned signals, often with insufficient principled backing. \emph{Learning-free methods} often exploit linguistic or statistical signals, such as stopword lists, IDF scores, or token positions within a document~\cite{acquaviaStaticPruningMultiRepresentation2023, liuAnalysisMatchingMechanisms2024}, retaining early tokens or content words with high IDF values while discarding later tokens or low-IDF tokens such as ''the'' or ''and''. These methods are simple and model-agnostic, making them easy to deploy across different retrieval architectures. However, they ignore the interaction between tokens and queries in the embedding space. A token that is \emph{statistically unimportant} may still be critical for distinguishing relevant from non-relevant documents for certain queries. In contrast, \emph{learning-based approaches} propose neural modules based on proxy signals to inform pruning decisions. For instance, AligneR~\cite{qianMultiVectorRetrievalSparse2022} proposed to employ entropy-regularized optimization (relaxed top-$k$ optimization), and ColBERTer~\cite{hofstatterIntroducingNeuralBag2022} reduces the number of vectors by combining sub-word vectors into unique whole-word representations and learns to discard them through a trainable gate vector. However, they provide limited theoretical backing, and very little work has focused on formalizing the pruning objective itself.

To our knowledge, \tllpaper~\cite{zongLosslessTokenPruning2025b} is the first to propose a formally defined lossless token pruning objective, aimed at removing only those tokens that do not affect retrieval scores. Although conceptually elegant, its experimental results demonstrate that achieving truly lossless pruning is extremely challenging in practice. Its experiments further show that applying this lossless objective in a lossy setting where a fixed subset of tokens must be retained can lead to substantial degradation in retrieval performance, particularly at very high pruning ratios. Additionally, its formulation requires document vectors to be of non-unit norms, further limiting its applicability. 

We address these limitations by introducing a principled and practical framework that formalizes the pruning objective in terms of expected retrieval error using hyperspace geometry. In this framework, we demonstrate that the importance of a token embedding can be accurately characterized by its Voronoi region~\cite{Aurenhammer1991VoronoiDS} in the embedding space, which represents the set of query tokens for which that token is most important. Building on this characterization, we derive a pruning algorithm that efficiently identifies tokens that can be pruned from a document while minimally impacting retrieval performance, by directly optimizing the pruning objective that other approaches attempt to indirectly approximate, resulting in pruning that minimizes retrieval degradation, even under aggressive pruning where as many as 90\% of the tokens in a document are removed. 

Beyond these improvements, our approach offers a new analytical perspective on token-level relevance, shedding light on the geometry of embedding spaces and their role in retrieval. 

Our main contributions are as follows: 
\begin{enumerate}
    \item We reframe \colbert{} token pruning as a Voronoi cell estimation problem and introduce a GPU-friendly pruning strategy based on this formulation that is highly effective, versatile, and up to 120 times faster than approximated LP-pruning~\cite{zongLosslessTokenPruning2025b}.
    \item We evaluate our pruning algorithm across varying pruning ratios, regularization strengths, and domains, demonstrating that Voronoi Pruning achieves state-of-the-art effectiveness among pruning methods, with particularly pronounced gains at aggressive pruning ratios where existing approaches degrade sharply.
    \item We highlight the broader utility of the formalism by using it to analyze empirical approaches such as first-$p$ token pruning and demonstrating the linear relationship between mean error (ME) and retrieval impact (nDCG@10), which can be used to guide pruning decisions.
    \nt{Maybe should emphasize more the correlation between MER and nDCG to contribution? It seems important in practice.}
\end{enumerate}

Collectively, our contributions establish a principled theoretical framework for token pruning while demonstrating its practical effectiveness in late-interaction retrieval. Our empirical results are instantiated within the \colbert framework, given its widespread adoption as a standard late-interaction architecture, but the proposed formulation depends only on the max-sim aggregation mechanism and the geometry of the embedding space, and therefore extends naturally to other late-interaction models.

The code for the pruning algorithm and associated experiments is publicly available on GitHub\footnote{\href{https://github.com/yash-reddy/voronoi-pruning}{https://github.com/yash-reddy/voronoi-pruning}}.

\section{Related Work}

\paragraph{Late-interaction Retrieval}

Late interaction models represent documents and queries as sets of token-level embeddings and compute relevance via fine-grained token interactions. With transformer-based architectures, late-interaction models rely on a simple interaction mechanism and achieve state-of-the-art performance. Models like \colbert~\cite{khattabColBERTEfficientEffective2020a, santhanamColBERTv2EffectiveEfficient2022a} or its variants \cite{gaoCOILRevisitExact2021,liCITADELConditionalToken2023,leeRethinkingRoleToken2023} determine query-document relevance by computing the highest inner product of the representation for each query token with that of a document token representation. Although such models allow building an index for first-stage retrieval, they still require much larger index sizes than single-vector models. To alleviate such problems, index compression and index pruning methods have been proposed. 

\paragraph{Index Compression}
A first approach to reduce the index size, which is orthogonal to our work, focuses on compressing embeddings rather than pruning them. Models such as \colbert and ColBERTer~\cite{hofstatterIntroducingNeuralBag2022} reduce the dimensionality of token embeddings directly using linear projection layers. \colbertvtwo~\cite{santhanamColBERTv2EffectiveEfficient2022a}, CRISP~\cite{venerosoCRISPClusteringMultiVector2025} and ~\cite{clavieReducingFootprintMultiVector2024}  cluster token representations to either fully replace token embeddings or significantly reduce their size. MUVERA~\cite{dhulipalaMUVERAMultiVectorRetrieval2024b} similarly compresses token representations into fixed-dimensional encodings that approximate multi-vector similarity. On a related note, XTR~\cite{leeRethinkingRoleToken2023} and PLAID~\cite{santhanamPLAIDEfficientEngine2022} also optimize retrieval latency by filtering token clusters using query-cluster dot products, using similar geometric constraints as clustering approaches. 

Although index compression reduces vector size, it retains redundant vectors. Index Pruning can remove these redundant, low-utility vectors, cutting storage and compute costs. 

\paragraph{Index Pruning}  
Past pruning approaches can broadly be classified as learning-free or learned methods. Learning-free methods apply heuristic criteria such as removing stopwords~\cite{acquaviaStaticPruningMultiRepresentation2023}, low-IDF tokens~\cite{acquaviaStaticPruningMultiRepresentation2023,liuAnalysisMatchingMechanisms2024}, retaining first-$p$ tokens, or attention-based proxies~\cite{liuAnalysisMatchingMechanisms2024, lassanceLearnedTokenPruning2022}. These approaches do not require additional training, enabling straightforward, model-agnostic pruning. Learning-based methods, on the other hand, employ auxiliary modules to identify low-utility tokens. ConstBERT~\cite{macavaneyEfficientConstantSpaceMultivector2025} uses a fully connected layer to prune a document into a fixed set of embeddings, while LeapMV~\cite{heTokenPruningOptimization2025} applies a fully connected layer with a Gumbel-softmax gate to produce discrete token-level pruning decisions. AligneR~\cite{qianMultiVectorRetrievalSparse2022} further frames multi-vector retrieval as a sparse alignment problem, jointly learning token salience and pairwise document–query interactions. More closely related to our work, \citep{zongLosslessTokenPruning2025b} prunes tokens that do not affect the similarity score, framing their identification as a linear programming (LP) problem. However, LP-Pruning is constrained by the need for a destructive dimensionality reduction to achieve a competitive pruning ratio, and is computationally demanding, making the method inapplicable for large document collections. By strategically relaxing the pruning objective, we show that our formulation avoids both of these constraints while yielding a substantially better effectiveness-efficiency trade-off across the pruning approaches discussed above.

\section{Problem Formulation}
\colbert models the relevance of a document $D$ for a given query $Q$ by first encoding tokens, producing a set of contextualized embeddings at the token-level. Let $D = \{d_1,\dots,d_m\}$ and $Q = \{q_1,\dots,q_l\}$ denote the document and query token embeddings, respectively. The relevance score is then computed by comparing each query token to all document tokens as follows:
\begin{equation}
\mathrm{\colbert}(Q,D)=\sum_{q \in Q}\max_{d \in D} q \cdot d
\end{equation}

While query and document vectors are normalized by \colbert \cite{khattabColBERTEfficientEffective2020a} to lie \emph{on} an $n$-dimensional unit sphere, other works, such as ~\citep{zongLosslessTokenPruning2025b} project them so that they lie \emph{within} the $n$-dimensional unit ball $\mathbb{B}^n$. 

To formulate the pruning objective, we build on the recent work presented in \citep{zongLosslessTokenPruning2025b}, which defines pruning as the process of removing tokens that do not affect retrieval scores, \emph{i.e.} those that never return the maximum inner product for any query $q$. In their setup, the goal of token pruning is to find $ {D'}\subset {D}$ such that

\begin{equation}
\label{eq: original-objective}
    \forall  q \in \mathbb{B}^n, \quad \max_{ d \in  D} \  q \cdot  d = \max_{ d \in  D'} \  q \cdot  d,
\end{equation}


\noindent in other words, $D'$ is obtained from $D$ by removing tokens that do not maximize the inner product of any query $q$ or equivalently, which are never solutions of $\operatorname{argmax}_d q \cdot d$.  

\tllpaper~\citep{zongLosslessTokenPruning2025b} estimates $D'$ by showing that a document vector $d$ never contributes to the maximum dot product for any given query, 
casting Equation~\ref{eq: original-objective} as a linear programming problem. However, identifying such a subset that perfectly preserves scores for all $q \in \mathbb{B}^n$ is challenging in practice, as most document tokens contribute to the maximum score for at least some query vectors, even if only in small regions. Consequently, the pruning objective described in Equation~\ref{eq: original-objective} is only effective when reducing the dimensionality of document vectors. \customnote{For example, LP pruning requires the vectors to be projected down to 20 dimensions from the original 128 to achieve competitive pruning performance.}

\customnote{Add an example of cases where this fails?}
\yuxuannote{Maybe it is too hard here. But we can say that LP pruning requires to truncate the vector dimension to less than 20 (in contrast with the original representation of 128 dimension) to achieve a decent pruning rate. Such reduction could tend to prune the representation that are actually important to IR performance. (And our final experimental result validate it, as a extremely small pruning rate, voronoi is much better than LP).  }

Given the inherent difficulty of achieving exact lossless pruning in practice, we propose a relaxation of the pruning objective in Equation~\ref{eq: original-objective} to a more practical and flexible alternative: finding a vector subset of a pre-determined size $k$ that minimizes the error introduced by pruning:
\begin{equation}
\label{eq:error-definition}
  D'_k = \underset{T \in \mathcal{P}_k(D)}{\operatorname{argmin}} 
  \;
  \mathbb{E}_{q \in \mathbb{B}^n}
  \!\left[ \max_{d \in D} q \cdot d - \max_{t \in T} q \cdot t
  \right]
\end{equation}

\begin{equation}
  \text{where} \quad 
  \mathcal{P}_k(D) = 
  \left\{ 
    T \subseteq D ;  |T| = k 
  \right\}.
\end{equation}

In the following sections, we demonstrate that this objective can be compactly \jlrannote{what does efficiently means here? compactly?}formulated as a Voronoi Cell pruning problem and illustrate how it can be approximated to develop a theoretically robust and effective method of pruning tokens.

\section{Methodology} \label{sec:method}
\subsection{Preliminaries}
Given a set of document token vectors $D = \{d_1, d_2, \dots, d_m\}$, we first define the \emph{Voronoi cell} of a document vector $d_i$ as the set of query vectors for which $d_i$ is the top-scoring match:

\begin{equation}
\mathcal{V}_i = \{ q \in \mathbb{B}^n : d_i = \arg\max_{d \in D} q \cdot d \}.
\end{equation}

The Voronoi cells partition the query space according to which document vector achieves the maximum dot product for each query.
We define the error introduced by pruning a single vector $d_i$ as the expected decrease in maximum similarity over all queries:

\begin{equation}
\label{eq:error-integral}
\begin{aligned}
\mathrm{Error}(d_i)
&= \mathbb{E}_{q \in \mathbb{B}^n}
  \left[
    \max_{d \in D} q \cdot d
    - \max_{d \in D \setminus \{d_i\}} q \cdot d
  \right] \\
&= \int_{\mathbb{B}^n}
  \left(
    \max_{d \in D} q \cdot d
    - \max_{d \in D \setminus \{d_i\}} q \cdot d
  \right)
  p_{\mathbb{B}^n}(q)\, dq
\end{aligned}
\end{equation}
where $p_{\mathbb{B}^n}(q)$ is the distribution over query token vectors.

\jlrannote{Remark that LP pruning amounts to remove tokens with empty Voronoi cells, so we can claim that our method is an extension/generalization of LP Pruning.} \yuxuannote{totally agree with it.}

Since removing $d_i$ only affects queries in its Voronoi cell $\mathcal{V}_i$, the integral reduces to $\mathcal{V}_i$. Moreover, decomposing $q = \alpha \hat{q}$ with $\hat{q} \in \mathbb{S}^n$ (the $n$-dimensional unit sphere\footnote{$\mathbb{B}^n = \{x \in \mathbb{R}^n : \|x\| \leq 1\}$ denotes the unit ball and $\mathbb{S}^n = \{x \in \mathbb{R}^n : \|x\| = 1\}$ its boundary, the unit sphere.}) and $\alpha \in [0,1]$ shows that cell membership and the $\operatorname{argmax}$ depend only on direction, so the radial component integrates out: 
\begin{align}
    \text{Error}(d_i)
    &= \int_{\mathcal{V}_i} \left( q \cdot d_i 
       - \max_{d \in D \setminus \{d_i\}} q \cdot d \right) 
       p_{\mathbb{B}^n}(q) \, dq \nonumber \\
    &= \int_{\mathcal{V}_i \cap \mathbb{S}^n} \int_0^1 
       \left( \alpha \hat{q} \cdot d_i 
       - \max_{d \in D \setminus \{d_i\}} \alpha \hat{q} \cdot d 
       \right) p(\alpha) \, d\alpha \, 
       p_{\mathbb{S}^n}(\hat{q}) \, d\hat{q} \nonumber \\
    &= \frac{n}{n+1} \int_{\mathcal{V}_i \cap \mathbb{S}^n} 
       \left( \hat{q} \cdot d_i 
       - \max_{d \in D \setminus \{d_i\}} \hat{q} \cdot d \right) 
       p_{\mathbb{S}^n}(\hat{q}) \, d\hat{q} \label{eq:error-sphere}
\end{align}

where $p_{\mathbb{S}^n}(q) = p(q \mid q \in \mathbb{S}^n)$ and $p(\alpha) = n\alpha^{n-1}$ is the radial density under the uniform distribution on $\mathbb{B}^n$. Since $\frac{n}{n+1}$ is constant across all tokens in a document, it doesn't affect relative rankings and is omitted in practice. The resulting error is governed solely by the gap between the first and second-best matches for each query in $\mathcal{V}_i$, making it both interpretable and efficient to compute. We note that if pruning is restricted to removing only those tokens whose pruning error is exactly zero, this naturally reduces to lossless pruning. Throughout this paper, we refer to the per-token quantity $\mathrm{Error}(d_i)$ as the \emph{pruning error} of token $d_i$, and to the average pruning error induced across a document (or corpus) at a given pruning ratio as the \emph{Mean Error} (ME).




\subsection{Voronoi Pruning Algorithm}\label{sec:vpa}
Having established a principled measure of token importance, we now outline \emph{Voronoi Pruning}, the algorithmic procedure that takes advantage of this measure to produce optimally pruned token subsets. Our method consists of four main components: 
\begin{enumerate}
    \item Monte Carlo Estimation of Expected Error
    \item Iterative token removal with error updates
    \item Global application across documents
    \item Search strategies to improve greedy selection.
\end{enumerate}

\paragraph{Monte Carlo Estimation of Expected Error}
Direct computation of the expectation in Equation~\ref{eq:error-integral} is intractable, as it requires integrating over the entire unit sphere. We therefore approximate the expectation via Monte Carlo sampling, assuming a uniform distribution of queries. We sample $N$ query vectors $Q_N = \{q_1,\dots,q_N\} \sim \mathcal{U}(\mathbb{S}^n)$ and calculate the expected pruning error for token $d_i$ as:
\begin{equation}
\label{eq:mc-error}
    \widehat{\mathbb{E}}[\mathrm{Error}(d_i)] = 
    \frac{1}{N}\sum_{q \in Q_N \cap \mathcal{V}_i} 
    \Bigl(q \cdot d_i - \max_{d \in D \setminus \{d_i\}} q \cdot d\Bigr).
\end{equation}

Following Equation~\ref{eq:error-sphere}, we restrict Monte Carlo sampling to unit-norm queries here, preserving the induced maximum dot-product partition while allowing for an efficient and stable estimation of mean error.


The choice of a uniform prior over $\mathbb{S}^n$ is the maximum-entropy distribution in the absence of a known query distribution: it assigns equal probability to all query directions, encoding no prior preference for any region of the embedding space. In practice, query distributions are rarely available, as evidenced by the lack of a train split of queries in 6 of the 13 publicly available BEIR datasets, and fitting a dataset-specific distribution would introduce dependence on specific query subsets and undermine the post-hoc, corpus-agnostic character of our method. The uniform prior avoids these failure modes while remaining well-grounded empirically. 

\begin{figure}[t]
\centering
\begin{subfigure}[t]{0.45\textwidth}
\centering
\includegraphics[width=\textwidth]{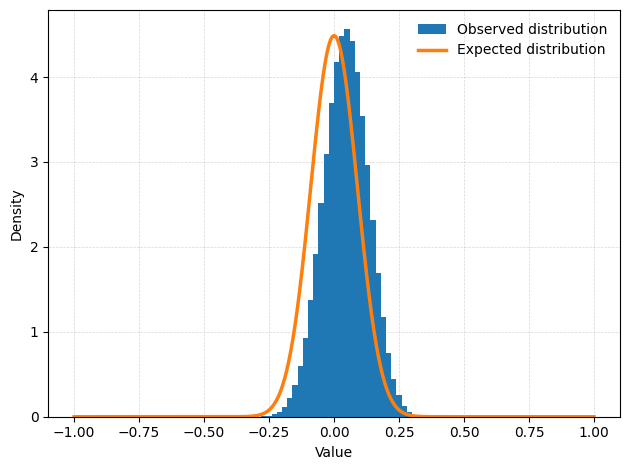}
\caption{Observed query embedding distribution along the most divergent dimension (dimension 64) from the expected uniform marginal.}
\label{fig:qdim-spread}
\end{subfigure}
\hfill
\begin{subfigure}[t]{0.45\textwidth}
\centering
\includegraphics[width=\textwidth]{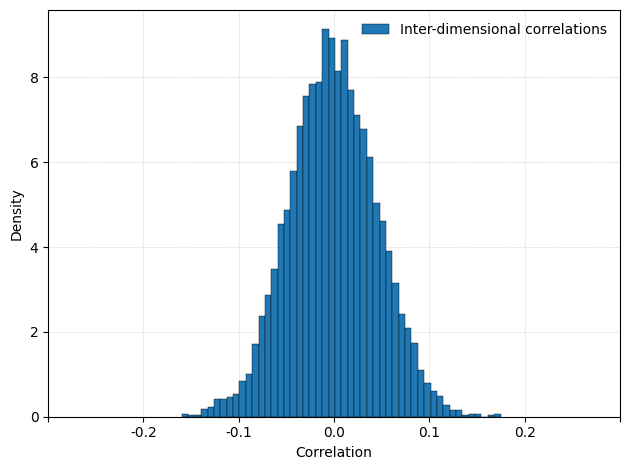}
\caption{Pairwise correlations between embedding dimensions.}
\label{fig:bin-corr}
\end{subfigure}
\caption{Embedding analysis of 100{,}000 MS~MARCO training queries encoded using \colbert.}
\label{fig:two-figs}
\end{figure}

\begin{figure}[t]
\centering
\begin{subfigure}[t]{0.45\textwidth}
\centering
\includegraphics[width=\textwidth]{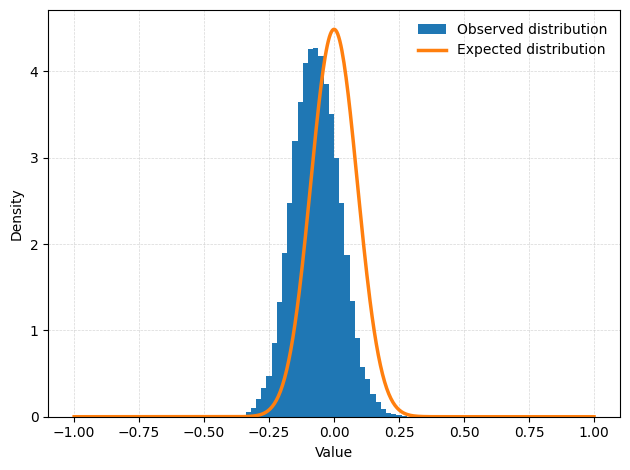}
\caption{Observed query embedding distribution along the most divergent dimension (dimension 83) from the expected uniform marginal.}
\label{fig:nf-qdim}
\end{subfigure}
\hfill
\begin{subfigure}[t]{0.45\textwidth}
\centering
\includegraphics[width=\textwidth]{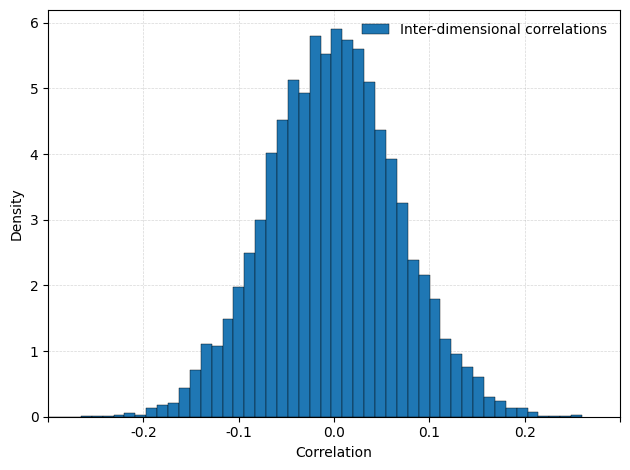}
\caption{Pairwise correlations between embedding dimensions.}
\label{fig:nf-corr}
\end{subfigure}
\caption{Embedding analysis of the 2{,}590 NFCorpus training queries encoded using \colbert.}
\label{fig:two-figs-nf}
\end{figure}

Empirically, this assumption holds across diverse query sets as well. While recent work has shown that transformer token embeddings are anisotropic within a single text~\cite{godeyAnisotropyInherentSelfAttention2024}, the distribution of embeddings across multiple queries remains approximately uniform. Analyzing 100k queries sampled from the MS~MARCO train set, alongside the train queries provided by SciFact, FiQA, and NFCorpus, we find that the worst-case per-dimension KL divergences from the theoretical uniform marginal, given by $p(x)= (1-x^2)^{62.5}$, are $0.13$, $0.19$, $0.20$, and $0.30$ respectively, illustrated in Figures~\ref{fig:qdim-spread} and~\ref{fig:nf-qdim} (for MS~MARCO and NFCorpus). Absolute pairwise covariances also fall predominantly in $[0, 0.15]$ with no dimension pairs exhibiting strong dependence, as shown in Figures~\ref{fig:bin-corr} and~\ref{fig:nf-corr}. These deviations introduce some imprecision in our Monte Carlo estimates, particularly in more specialized out-of-domain settings such as NFCorpus, but remain sufficiently small to constitute a manageable approximation in practice, an assumption further corroborated by the strong out-of-domain retrieval performance reported in Section~\ref{sec:results}.

Estimating theoretical confidence intervals for $\mathrm{Error}(d_i)$ is challenging, as it follows a compound distribution arising from a two-stage sampling process: (i) a Bernoulli distribution capturing the probability $p$ that a sampled query embedding falls within a given Voronoi cell, and (ii) the conditional distribution of error values $\mathrm{Error}(d_i)$ observed within that cell, which is non-standard and generally unknown. However, the first component of this process is well-behaved. Estimating confidence intervals (CI) for the probability of cell membership $p$ is straightforward since membership, even in high-dimensional polytopes, is a Bernoulli variable. The widest CIs occur for $\hat{p} = 0.5$, and with just $10^4$ sampled queries, a normal approximation of $\hat{p} = 0.5$ yields a $95\%$ CI of $[0.4902, 0.5098]$, corresponding to an absolute uncertainty of less than $1\%$. Although the conditional error distribution remains difficult to characterize theoretically, the ability to accurately estimate cell membership ensures that the dominant source of uncertainty in our Monte Carlo procedure is well controlled, making the resulting pruning estimates empirically stable and useful. Finally, we note that sampling $10^4$ vectors, as well as computing their inner products with document token representations, is much more efficient, and stable, than following the approximate procedure in~\citep{zongLosslessTokenPruning2025b}, as demonstrated in our experiments.

\begin{figure*}[htbp]  
    \centering
    \begin{subfigure}[b]{0.3\linewidth}
        \centering
        \includegraphics[width=\linewidth]{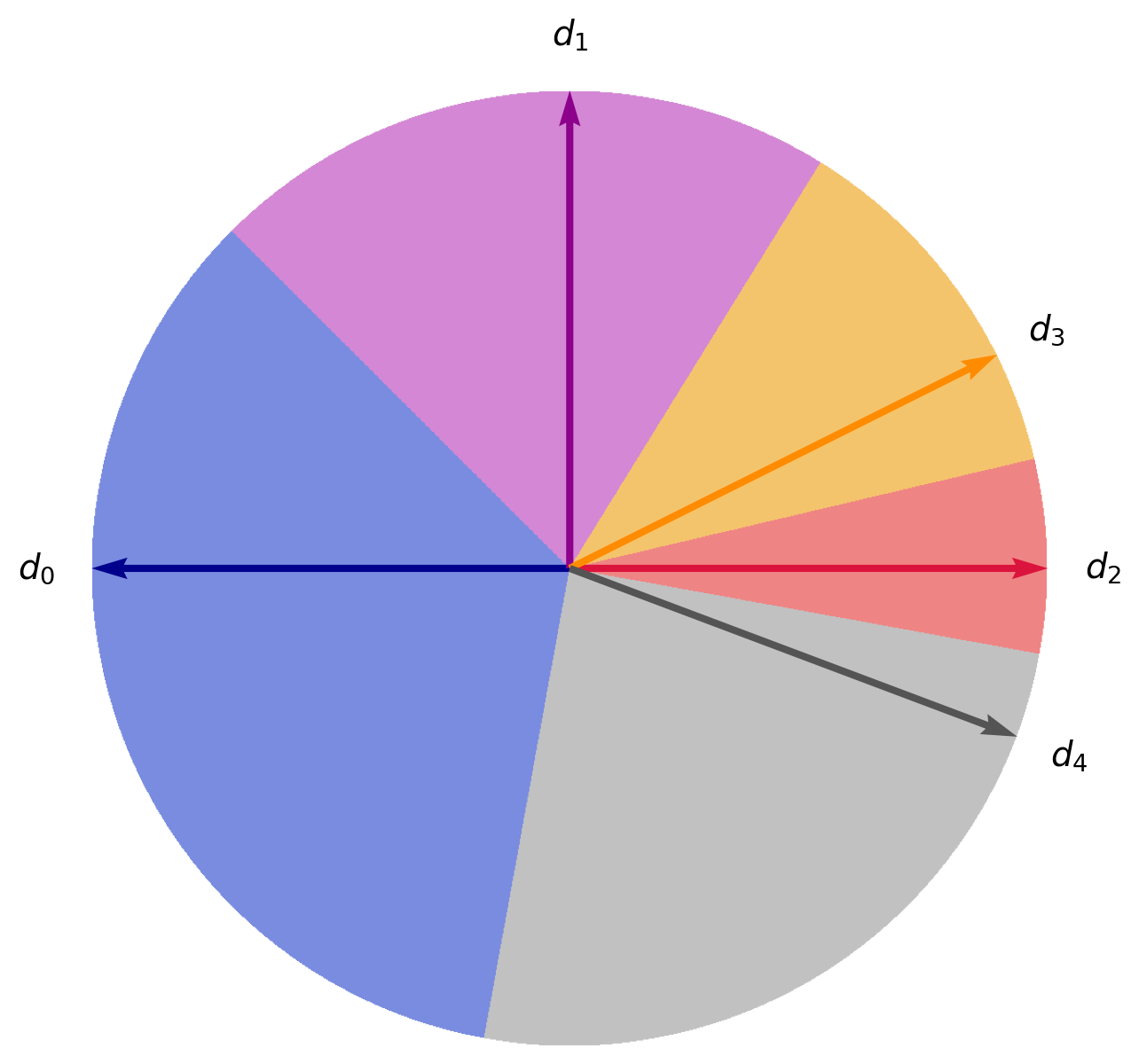}
        \caption{Initial configuration consisting of 5 document vectors.}
        \label{fig:initial}
    \end{subfigure}
    \hfill
    \begin{subfigure}[b]{0.28\linewidth}
        \centering
        \includegraphics[width=\linewidth]{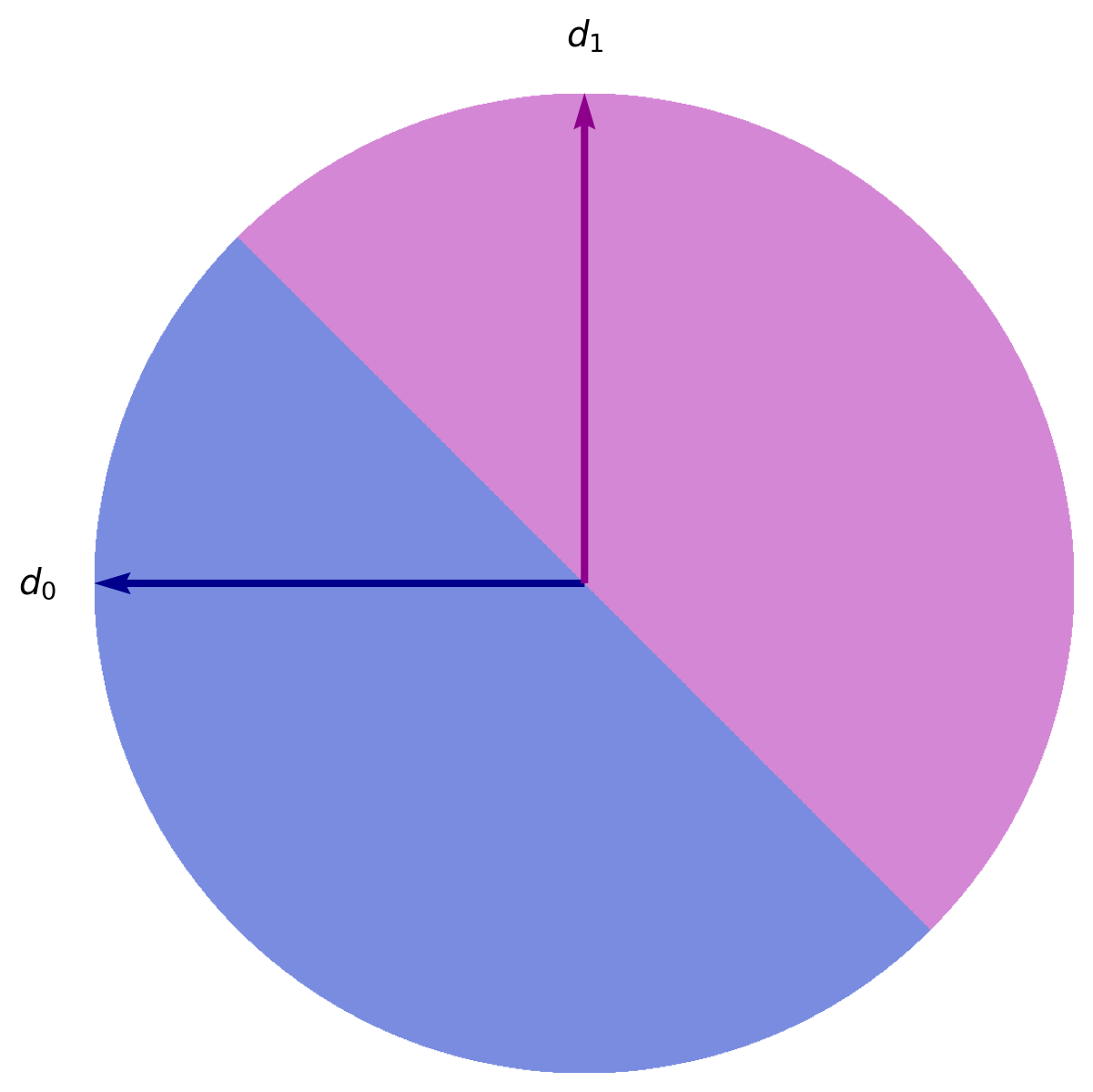}
        \caption{Non-iterative pruning: vectors $d_0$ and $d_1$ retained; Mean error: 0.3872.}
        \label{fig:noniterative}
    \end{subfigure}
    \hfill
    \begin{subfigure}[b]{0.3\linewidth}
        \centering
        \includegraphics[width=\linewidth]{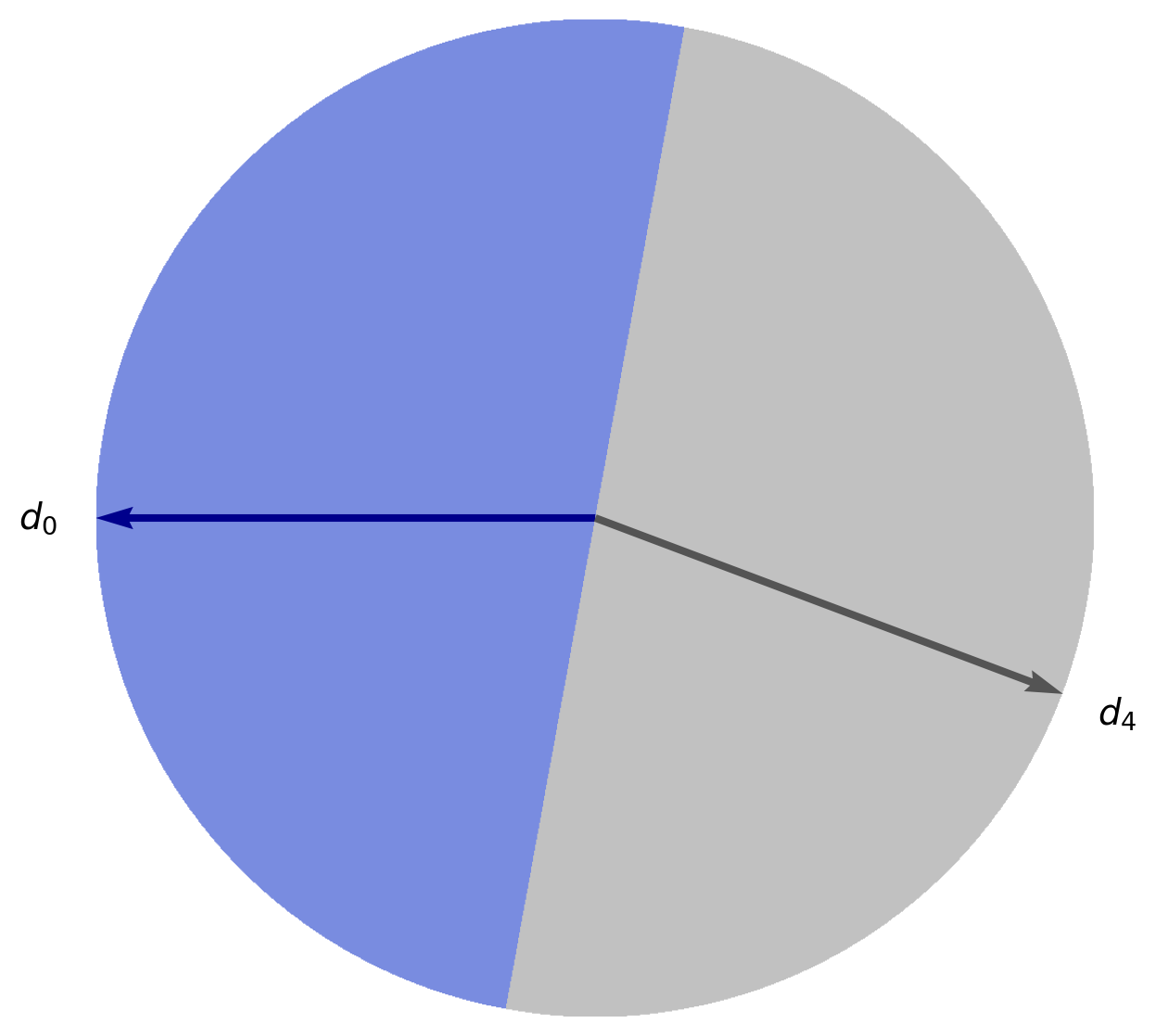}
        \caption{Iterative pruning: vectors $d_0$ and $d_4$ retained; Mean error: 0.2100.}
        \label{fig:iterative}
    \end{subfigure}
    \caption{An example illustrating the difference in iterative and non-iterative Voronoi pruning of 2D document vectors. Each subfigure shows the maximum dot-product Voronoi regions and the retained document vectors.}
    \label{fig:pruning_comparison}
\end{figure*}

\paragraph{Iterative Pruning}
Pruning is inherently structural, and removing a token reshapes the surrounding Voronoi cells, altering the contribution of neighboring tokens. A one-shot strategy that evaluates token errors only once may, therefore, yield suboptimal subsets.
To address this, we adopt an \emph{iterative pruning scheme}. At each iteration, the token with the smallest current pruning error is removed and the error estimates of all remaining tokens are recomputed under the updated Voronoi diagram. This adaptive update ensures that subsequent decisions reflect the evolving structure of the representation space and leads to higher-quality pruned sets. The advantage of this strategy is illustrated in Figure~\ref{fig:pruning_comparison}, where iterative pruning (\autoref{fig:iterative}) preserves a lower mean error in the maximum dot-product values compared to non-iterative pruning.

The iterative nature of our algorithm is practically manageable for two reasons. Firstly, when a document token is pruned, only the queries previously assigned to its Voronoi cell need to be reassigned; the membership of other query vectors remains unchanged. Secondly, most of the computations involved rely on dot-product operations, which are highly optimized on modern GPUs. Consequently, although iterative updates introduce considerable computational overhead, an efficient implementation allows the approach to scale easily to large datasets with minimal difficulty.

\paragraph{Global Pruning}
Because our pruning error is consistently defined across all documents, we extend the procedure to operate at the \emph{collection-level}.
Instead of pruning independently within each document, tokens are ranked globally by their error contributions, and the lowest-impact token is removed at each step until a target budget (e.g., a fixed number or fraction of tokens) is reached. This allows for more efficient identification and pruning of low-importance tokens at the \emph{collection-level}.

In practice, we pre-compute document-level pruning orders in parallel and merge them according to the pruning errors they introduce to obtain the global ranking.

\paragraph{Improving Greedy Pruning}
Finding the globally optimal subset of tokens would require evaluating all possible token subsets, a computationally intractable \jlrannote{\emph{intractable} is more appropriate. Feasible meaning is overloaded for optimisation.}\customnote{Y: Makes sense, I've replaced feasible with intractable here} combinatorial problem. Although our iterative procedure provides a tractable greedy approximation, we note that it optimizes for local rather than global optimality. To partially mitigate this limitation, we explore using \emph{beam search} inspired by decoding strategies in language modeling. During pruning, multiple candidate subsets are evaluated at each step, and only the top-$k$ beams with the lowest total pruning error are retained for further iteration. This encourages the exploration of higher-quality token subsets without incurring the full combinatorial cost.

Algorithm~\ref{algo:simple-vp} describes a simplified version of our iterative document-level pruning, without beam search.

\begin{algorithm}[tb]
\DontPrintSemicolon
\SetAlgoLined
\SetKwInOut{Input}{Input}
\SetKwInOut{Output}{Output}
\SetKwInOut{StopCondition}{Stop Condition}
\caption{Simplified Single-Document Voronoi Pruning}\label{algo:simple-vp}

\Input{Token vectors $D = \{d_1, \dots, d_m\}$; sample points $Q_N = \{q_1, \dots, q_N\}$; Target size $t$}
\Output{Pruned token set $D' \subset D$ with $|D'| = t \leq m$}

\ForEach{$d \in D$}{
    Compute $\varepsilon_d= \mathrm{Error}(d,Q_N,D)$ using Equation~\ref{eq:mc-error}\;
}

\While{$|D| > t$}{
    $d_{\min} = \arg\min_{d \in D} \varepsilon_d$\;
    $D = D \setminus \{d_{\min}\}$\;
    
    \ForEach{$d \in D$}{
         Recompute $\varepsilon_d$ using Equation~\ref{eq:mc-error}\;
    }
}

\Return{Pruned token set $D$}
\end{algorithm}

\section{Experimental Setup}
\subsection{Implementation Details}
Our experimental setup follows the previously established framework from \citep{zongLosslessTokenPruning2025b}. All pruning experiments are conducted starting with the pre-trained \colbertvtwo{} checkpoints\trimbin{unless specified otherwise}. When comparing against fine-tuned models, we adopt the same two-stage re-ranking pipeline, using SPLADEv2 as the first-stage retriever and \colbertvtwo{} as the reranker, to ensure a direct comparison. Following their setup, we also apply a projection layer that maps document token vectors from unit norm to norms in $[0,1]$ for all fine-tuned checkpoints. We utilize two of the strongest propositions from \citep{zongLosslessTokenPruning2025b} in terms of effectiveness and efficiency trade-off: the L1 norm regularization ($\mathcal{L}^{L1}$), which minimizes the norm of the document token representation vectors: 
\begin{equation}
    \mathcal{L}^{(L1)} = \frac{1}{m}\sum_{d\in D} \|d\|_1
\end{equation}
and the Doc-sim regularization: 
\begin{equation}
    \mathcal{L}^{(sim)} = -\frac{1}{m(m-1)}\sum_{d\in D}\left(1 - \|d\|_2\right)\sum_{d'\in D\setminus\{d\}} \frac{[d\cdot d']_+}{\|d\|_2 + \varepsilon}
\end{equation}
where $m$ denotes the number of document token vectors across a document and $\varepsilon=0.01$ to avoid dividing by 0. In our experiments, the strength of the regularization is controlled by the coefficient $\alpha$. To simplify the notation, we refer to \colbertvtwo{} fine-tuned with the doc-sim and L1 regularizers as \colbertvtwo{-sim} and \colbertvtwo{-L1} respectively, abbreviated as $\alpha \mathcal{L}^{(sim)}$ and $\alpha \mathcal{L}^{(L1)}$ in tables, with regularization coefficients noted in parentheses. Unless otherwise specified, we evaluate both as second-stage rerankers using SPLADEv2 as the first-stage retriever.

To approximate the expected pruning error defined in Section~\ref{sec:method}, we use $10^5$ sample points when working with the full \colbertvtwo{} index and $10^4$ points during two-stage retrieval experiments, balancing computational costs and accuracy.

As discussed later in Section~\ref{sec:results}, we find that beam search is impractical in our setting, even with a beam size of three. It incurs substantial memory overhead and leads to significant computational slowdowns, while yielding no measurable improvement in pruning order or downstream performance. Accordingly, all experiments in this paper use the full version of our algorithm incorporating all proposed components except beam search, which we refer to as \emph{Voronoi Pruning}. As such, beam search is evaluated only in our ablation studies and is omitted from all other experiments.

All experiments were run on either NVIDIA V100 or H200 GPUs, depending on availability.

\subsection{Datasets}

The in-domain accuracy of our Voronoi Pruning (VP) approach is evaluated using the MS~MARCO v1 passage retrieval task~\cite{bajaj2016ms}. The task consists of 8.8M passages paired with real user queries and relevance annotations, and serves as our primary benchmark for evaluating pruning effectiveness. We conduct our evaluation on the MS~MARCO dev set (6,980 queries) as well as the 2019 (43 assessed topics) and 2020 (54 assessed topics) iterations of the TREC Deep Learning track~\cite{craswell2019overview, craswell2020overview}, which also utilize the 8.8M passage corpora from MS~MARCO. To assess robustness under domain shift, we evaluate in a zero-shot setting on the same subset of the BEIR benchmark~\cite{thakur2021beir} as~\cite{zongLosslessTokenPruning2025b}, which excludes HotpotQA, FEVER, and ArguAna, as the longer documents in these collections make LP-Pruning prohibitively expensive, precluding a direct comparison. We report MRR@10 for the MS~MARCO dev set, and nDCG@10 for TREC-DL and BEIR.

\subsection{Baselines}
We compare our methodology against a diverse set of baselines that span both traditional information retrieval approaches and recent state-of-the-art developments. 
Specifically, we include:
\paragraph{\textbf{Retrieval Baselines}} - We compare our approach against three widely used, unpruned retrieval baselines. BM25 is a classical lexical matching baseline and remains a widely adopted reference in IR research due to its simplicity and strong empirical performance. SPLADEv2~\cite{formalSPLADEV2Sparse2021} is a state-of-the-art learned sparse retriever that is also used as a first-stage retriever in our re-ranking experiments.  We also include the performance metrics of \colbertvtwo~\cite{santhanamColBERTv2EffectiveEfficient2022a} before our pruning algorithm and fine-tuning routine are applied, in both its end-to-end configuration and re-ranking setup. In the end-to-end setting (\colbertee), we use the original configuration of \colbertvtwo, building an index and retrieving the documents in a single stage using PLAID~\cite{santhanamPLAIDEfficientEngine2022}. In the re-ranking setup (\colbertrr), we re-rank the top 100 documents retrieved by SPLADEv2 using \colbertvtwo. This enables a direct comparison with late-interaction retrieval performance and quantifies the extent to which effectiveness is preserved under various scoring and pruning strategies.
    
\paragraph{\textbf{Static Pruning Strategies}} - We compare against four heuristic-based methods for token pruning: stopword removal, low-IDF pruning, positional pruning, and attention score-based pruning.
    
\paragraph{\textbf{Learned Pruning Strategies}} - We further compare against learned pruning approaches, which dynamically identify and reduce low-importance tokens through additional modules that typically require fine-tuning. Approaches such as AligneR~\cite{qianMultiVectorRetrievalSparse2022} attempt to predict pruning decisions on a token-level to decide which tokens to retain, whereas ConstBERT~\cite{macavaneyEfficientConstantSpaceMultivector2025} aims to represent document tokens using a fixed number of embeddings. ColBERTer~\cite{hofstatterIntroducingNeuralBag2022} is another such learned pruning approach that proposes to reduce the size of token embeddings while retaining the original size of $\left[\mathtt{CLS}\right]$ token embeddings, resulting in an overall reduction of the index size. We also compare against the dominance-based pruning and regularization framework introduced by~\cite{zongLosslessTokenPruning2025b}, which formulates pruning as the elimination of token embeddings that are redundant under vector dominance constraints. In particular, we report comparisons against two of the strongest pruning strategies introduced in that work, namely \emph{Norm Pruning} (NP), which prunes the document token which has a norm smaller than a threshold $\theta = 0.5$, and \emph{LP-Pruning} (LPP), which prunes the document token that diagnostically has a low contribution to the query-document token interaction, with a threshold $\theta=0.7$.   


The selected baselines are highly diverse, motivated by different paradigms (including linguistic, mathematical, and neural approaches) and differ widely in computational costs, fine-tuning requirements, and overall versatility. By evaluating our proposed approach against this set of baselines, we present a comprehensive understanding of its strengths and limitations.

\section{Results and Analysis}\label{sec:results}
\subsection{Main Results}
\subsubsection{In-domain retrieval} Table~\ref{tab:MS-MARCO-perf} reports MRR@10 on the MS~MARCO dev set under varying pruning strategies. Among all learning-free methods, Voronoi Pruning achieves the highest effectiveness while maintaining comparable pruning ratios, preserving \textbf{98.0\%} of the unpruned \colbertee{} performance at a \textbf{50\%} token budget and convincingly outperforming heuristic baselines such as IDF-based, positional, and attention-score pruning.

Voronoi Pruning approaches the performance of other learned pruning methods, such as AligneR, at moderate pruning ratios, despite operating entirely post hoc without additional fine-tuning or architectural modifications. When applied to \colbertvtwo{-sim} and \colbertvtwo{-L1} checkpoints, Voronoi Pruning also competes with the strongest learned pruning strategies, including Norm-Pruning and LP-Pruning, matching their effectiveness. However, unlike Norm-Pruning and LP-Pruning, our method can also be applied directly to embeddings with unit norm, increasing its applicability across different pre-trained models.


Moreover, we note that Voronoi Pruning achieves this competitive performance while being substantially faster than LP-Pruning. On 10,000 sampled MS~MARCO documents, Voronoi Pruning with $10^4$ randomly sampled query vectors requires 12.0 seconds on a single H200 GPU ($\approx$1.2ms per document), compared to 1,474.3 seconds for LP-Pruning on CPU ($\approx$147ms per document). This per-document gap holds regardless of parallelization strategy, as both methods distribute trivially across documents. In practice, this means that pruning the full 8.8M MS~MARCO collection requires approximately 3 hours, whereas LP-Pruning would take an estimated 15 days. This difference is particularly relevant for dynamic indexes, where documents are regularly added or removed and pruning becomes a recurring cost rather than a one-time expense.

\begin{table}[t]
\centering
\caption{MRR@10 and remaining token percentages for various pruning strategies on the MS~MARCO dev set. VP denotes our Voronoi Pruning method.} 
\resizebox{\columnwidth}{!}{
\begin{tabular}{@{}lcccc@{}}
\Xhline{1pt}
\textbf{model}  & \textbf{Remain \%} & \textbf{Dev set} & \textbf{DL19} & \textbf{DL20} \\
\Xhline{1pt}
\multicolumn{2}{l}{\textit{Without Pruning}} \\
\hline
BM25 &  - & 18.7 & 50.6 & 47.5 \\
SPLADEv2  & - & 35.8 & 70.6 & 68.7 \\
\colbertee  & - & 39.7 & 74.5 & - \\
\colbertrr  & - & 40.0 & 74.4 & 75.6 \\
\Xhline{1pt}
\multicolumn{2}{l}{\textit{With Learning-free Pruning}} \\
\hline
First $p$ &  50\% & 37.7 & \textbf{72.3} & -\\
IDF  & 50\% & 32.6 & 70.2 & - \\
Stopwords  & 67\% & 30.9 & 70.9 & 67.8\\
Att. Score  & 50\% & 36.0 & 72.0 & -\\
\hline
Voronoi Pruning (Ours) &  50\% & \textbf{38.9} & 72.2 & \textbf{73.1}\\
\Xhline{1pt}
\multicolumn{2}{l}{\textit{With Learned Pruning}} \\
\hline
AligneR$_{\text{base}}$  & 40\% & 38.1 & - & -  \\
ConstBERT &  47\% & 39.0 & 73.1 & 73.3 \\
$\alpha \mathcal{L}^{(sim)}$($\alpha=0.8$) + LP-Pruning  & 32\% & \textbf{39.7} & \textbf{73.8} & \textbf{73.3}\\
$\alpha \mathcal{L}^{(sim)}$($\alpha=0.1$) + LP-Pruning  & 52\% & \textbf{39.9} & \textbf{74.2} & \textbf{73.8} \\
$\alpha \mathcal{L}^{(L1)}$($\alpha=0.01$) + Norm-Pruning  & 31\% & 39.2 & 73.6 & 72.1 \\
\hline
$\alpha \mathcal{L}^{(sim)}$($\alpha=0.8$) + VP (Ours) & 32\% & 39.5 & 73.7 & 73.0 \\

\Xhline{1pt}

\end{tabular}
}
\label{tab:MS-MARCO-perf}
\end{table}

\subsubsection{Out-of-domain Retrieval} Table~\ref{tab:beir-eval} summarizes zero-shot performance on the BEIR benchmark. Voronoi pruning consistently outperforms all learning-free baselines across the majority of tasks, achieving the highest average nDCG@10 among pruning methods that do not require retraining. Performance degradation remains modest even at aggressive pruning levels, indicating robustness under domain shift.

In several datasets (e.g., FiQA, NFCorpus, and TREC-COVID), Voronoi pruning achieves comparable or higher nDCG@10 scores relative to more expensive learned pruning approaches, while requiring no additional fine-tuning. These results suggest that a geometry-aware pruning objective offers competitive generalization across domains compared to heuristic or token-position-based strategies, which are often sensitive to dataset-specific token distributions.

Finally, when analyzing Voronoi Pruning against learned pruning baselines, it matches the performance of LP-Pruning and Norm-Pruning at equivalent pruning ratios, at a fraction of the computational cost.

\subsubsection{Performance degradation}
While evaluations at fixed token budgets provide a useful baseline, the ideal efficiency-effectiveness trade-off varies considerably across both experimental and applied contexts. To provide a comprehensive assessment of how retrieval performance degrades as pruning is performed more aggressively, we study the average nDCG@10 performance of different pruning methods on \colbertvtwo{-sim} ($\alpha=0.8$) embeddings for TREC-DL19 and TREC-DL20 datasets across multiple pruning ratios.

\begin{figure}
    \centering
    \includegraphics[width=\linewidth]{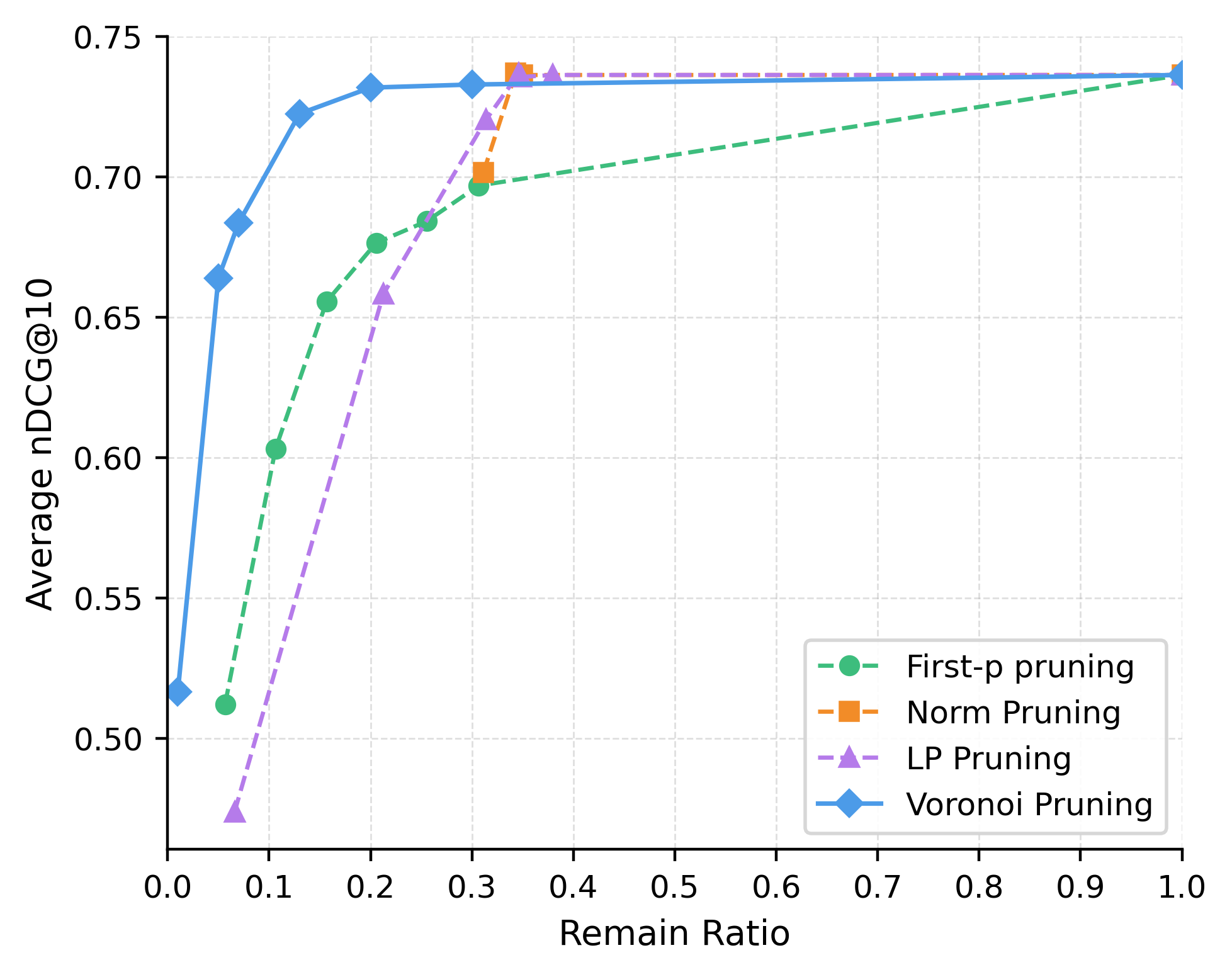}
    \caption{Average TREC DL (2019 and 2020) nDCG@10 for different pruning strategies on document vectors produced by \colbertvtwo{-sim} ($\alpha=0.8$).}
    \label{fig:prune-deg}
\end{figure}

Figure~\ref{fig:prune-deg} plots the performance degradation of different pruning methods against the fraction of tokens remaining after pruning. At moderate token budgets where about $30\%$ of the original tokens remain, the evaluated methods demonstrate broadly comparable effectiveness. However, as pruning becomes increasingly aggressive, a stark divergence emerges between Voronoi Pruning and its alternatives. For example, when the index is pruned to retain only \textbf{6\%} of its tokens, Voronoi Pruning successfully maintains an average nDCG@10 of \textbf{0.67}, whereas LP-Pruning falls sharply to \textbf{0.46} under the same conditions. This widening performance gap and Voronoi Pruning's ability to remain competitive well into extreme pruning scenarios further highlights the strength of its geometric grounding. By directly minimizing the expected retrieval error rather than relying on proxy approximations, our formulation produces a highly stable degradation profile that sustains retrieval quality far beyond the point where baseline approaches deteriorate.

\customnote{TODO: Decide what to do with the limitations}

\subsection{Ablation Studies}
We conduct ablation studies to understand the impact of individual design choices introduced in Section~\ref{sec:vpa}. While each component of the algorithm is theoretically motivated, it also introduces non-trivial computational overhead. These studies provide insights into how each design choice contributes to overall performance while also informing practical decisions about accuracy–efficiency trade-offs. We perform these experiments using a random sample of 100,000 MS~MARCO passages encoded using \colbert on a single V100 GPU and report the results in Table~\ref{tab:ablation-perf}.

\begin{table}[t]
\caption{Ablation study of pruning design choices on MS~MARCO passages, measuring MRR@10 performance of ColBERT embeddings (50\% pruned) on the MS~MARCO dev set and the time taken to generate pruning orders for 100k passages.}
\label{tab:ablation-perf}
\centering
\resizebox{\linewidth}{!}{
\begin{tabular}{lcr}
\toprule
Pruning Configuration & MRR@10 & Runtime (s) \\
\midrule
Voronoi Pruning (no ablation) & 38.9 & 602.55 \\
\midrule
Beam size = 3 & 38.9 & 3088.88 \\
Local pruning & 38.4 & 602.55 \\
Step size = 3 & 38.0 & 217.28\\
Non-iterative pruning & 33.2 & 28.97\\
\bottomrule
\end{tabular}
}
\end{table}

\paragraph{Effect of Beam Size} In principle, beam search could help bridge the gap between our greedy pruning strategy and the global optimum. In practice, however, using a beam size of 3 produced no improvement in MRR@10 while increasing runtime roughly fivefold. Given the marginal gains and substantial computational overhead, we find that the standard greedy approach offers the best trade-off between efficiency and effectiveness.
 
\paragraph{Global vs Local pruning} Pruning can be done on either the document level(local pruning) or the collection level (global pruning). In global pruning, pruning error scores are compared across documents, and tokens with the smallest scores are removed before updating the remaining scores. It must be noted that in both settings, the scores are still calculated at the document level, and pruning a token vector from a document only affects the scores of the tokens in that particular document. \textbf{Global pruning} achieves an MRR@10 score of \textbf{38.9} on the MS~MARCO dev set, compared to \textbf{38.4} for \textbf{local pruning}. Since pruning orders are generated at the document level, they only need to be merged once to facilitate global pruning. Consequently, there is very little difference between the execution times for global and local pruning.

\paragraph{Iterative vs Non-iterative pruning} Although the importance of dynamic pruning can be theoretically useful, it is a relatively compute-heavy process. As such, we investigated the performance gained by dynamically updating the scores after pruning each vector, against a pruning setup where only the initial pruning error scores were used for deciding pruning candidates, hereby referred to as the iterative and non-iterative pruning setups, respectively. We found that the \textbf{MRR@10} score drops steeply from \textbf{38.9} to \textbf{33.2} when replacing the iterative pruning setup with non-iterative pruning, highlighting the importance of dynamically updating pruning error scores throughout the pruning process.

\customnote{pruning error vs mean error - Need to check if this is confusing when reading the paper.}
\paragraph{Effect of Step Size}
Since recomputing pruning errors after the removal of each individual token is computationally expensive, yet fully non-iterative pruning leads to substantially degraded performance, we explore a compromise between the two extremes: rather than pruning a single token per iteration and recomputing errors each time, we prune multiple tokens at each step, thereby reducing the number of recomputations required. In experiments with a step size of 3, this strategy results in an almost threefold speedup, but also leads to a noticeable drop in effectiveness, with MRR@10 decreasing from \textbf{38.9} to \textbf{38.0}.

\customnote{Might want to say "except beam search" once the experiment is added.}In conclusion, we note that while each of our design decisions imposes a non-trivial computational overhead for Voronoi Pruning, except for beam search, each of these components has a considerable, positive effect on the performance of our proposed approach, and it is generally advisable to use our approach as originally described, without beam search, to achieve optimal performance.

\begin{table*}[ht]
\centering
\footnotesize 
\caption{nDCG@10 scores and remaining token percentage of different pruning strategies on the BEIR benchmarks. VP refers to Voronoi Pruning; *Indicates a deviation of our implementation's scores (provided in braces) from previously reported scores.}
\resizebox{\linewidth}{!}{
\begin{tabular}{m{2cm}m{1cm}C{1.2cm}C{0.8cm}C{0.8cm}C{0.8cm}C{0.8cm}C{0.8cm}C{0.8cm}C{0.8cm}C{0.8cm}C{0.8cm}C{0.8cm}C{0.8cm}C{0.8cm}C{0.8cm}|C{0.8cm}}
\Xhline{1pt}
\multicolumn{2}{c}{Model} & Remain \% & CF & DB & FQ & NF & NQ & QU & SD & SF & TC & TO & Avg. \\
\Xhline{1pt}
\multicolumn{14}{ l }{\emph{Without Pruning}} \\
\Xhline{0.5pt}
BM25 &  & - & 21.3 & 31.3 & 23.6 & 32.5 & 32.9 & 78.9 & 15.8 & 66.5 & 65.6 & 36.7 & 40.5 \\
SPLADEv2 &  & - & 21.3 & 42.0 & 31.6 & 32.8 & 50.8 & 18.1 & 14.0 & 65.9 & 65.5 & 25.5 & 43.0 \\
\colbertee &  & - & 17.6*\textcolor{gray}{(18.0)} & 44.6 & 35.6 & 33.8 & 56.2*\textcolor{gray}{(55.9)} & 85.2 & 15.4 & 69.3 & 73.8 & 26.3*\textcolor{gray}{(25.4)} & 45.8 \\
\colbertrr & & - & 20.3 & 45.9 & 35.7 & 34.3 & 56.3 & 85.7 & 14.4 & 68.5 & 75.5 & 33.7 & 47.0 \\
\Xhline{1pt}
\multicolumn{14}{l}{\emph{Learning-free Pruning}} \\
\Xhline{0.5pt}
First $p$ & & \textcolor{gray}{75\%} & 15.5 & 44.0 & 32.4 & 32.4 & 53.2 & 78.7 & 15.2 & 61.6 & 72.1 & 26.1 & 43.1 \\
IDF & & \textcolor{gray}{75\%} & 17.0 & \textbf{45.1} & 34.0 & 32.6 & 55.0  & 85.3 & \textbf{15.5} & 64.8 & 71.4 & \textbf{26.2} & 44.7 \\
Att. Score & & \textcolor{gray}{75\%} & 16.7 & 44.4 & 33.6 & 32.6 & 54.7 & 84.4 & \textbf{15.5} & 64.0 & 70.0 & \textbf{26.2} & 44.2 \\
VP (Ours) & & \textcolor{gray}{75\%} & \textbf{17.8} & 44.7 & \textbf{35.6} & \textbf{33.9} & \textbf{55.6} & \textbf{85.4} & \textbf{15.5} & \textbf{69.5} & \textbf{73.0} & 25.4 & \textbf{45.6} \\
VP (Ours)& & \textcolor{gray}{50\%} & 16.5 & 43.1 & 34.9 & 33.8 & 53.6 & 84.9 & 15.4 & 68.6 & 70.7 & 22.9 & 44.4 \\
\Xhline{1pt}
\multicolumn{14}{l}{Learned Pruning}\\ 
\Xhline{0.5pt}
\multirow{2}{*}{$\alpha \mathcal{L}^{(L1)}$ + LPP} & 
$\alpha = 0.01$ & - & 18.6 & 44.3 & 34.3 & $33.8$ & 54.0 & 84.5 & $14.2$ & $67.0$ & 71.6 & 31.8 & 45.4 \\
& Remain \% & - & \textcolor{gray}{34\%} & \textcolor{gray}{39\%} & \textcolor{gray}{30\%} & \textcolor{gray}{29\%} & \textcolor{gray}{35\%} & \textcolor{gray}{76\%} & \textcolor{gray}{32\%} & \textcolor{gray}{30\%} & \textcolor{gray}{32\%} & \textcolor{gray}{28\%} & \textcolor{gray}{39\%} \\
\hline
\multirow{4}{*}{$\alpha \mathcal{L}^{(sim)}$ + LPP} & $\alpha = 0.1$ & - & $20.2$ & 44.7 & $35.0$ & $34.4$ & 55.5 & 84.4 & 14.0 & $67.4$ & 71.1 & $32.8$ & 45.9 \\
& Remain \% & - & \textcolor{gray}{60\%} & \textcolor{gray}{58\%} & \textcolor{gray}{57\%} & \textcolor{gray}{64\%} & \textcolor{gray}{57\%} & \textcolor{gray}{58\%} & \textcolor{gray}{58\%} & \textcolor{gray}{69\%} & \textcolor{gray}{65\%} & \textcolor{gray}{65\%} & \textcolor{gray}{61\%} \\
\Xcline{2-14}{0.1pt}
& $\alpha = 0.8$ & - & 19.6 & \textbf{42.9} & \textbf{35.0} & \textbf{33.8} & \textbf{54.7} & \textbf{82.6} & 13.9 & \textbf{66.9} & 74.3 & 33.1 & \textbf{45.7} \\
& Remain \% & - & \textcolor{gray}{36\%} & \textcolor{gray}{34\%} & \textcolor{gray}{36\%} & \textcolor{gray}{38\%} & \textcolor{gray}{36\%} & \textcolor{gray}{40\%} & \textcolor{gray}{36\%} & \textcolor{gray}{38\%} & \textcolor{gray}{37\%} & \textcolor{gray}{40\%} & \textcolor{gray}{37\%} \\
\hline
\multirow{2}{*}{$\alpha \mathcal{L}^{(sim)}$ + VP (Ours)} & $\alpha=0.8$ & - & \textbf{19.9} & \textbf{42.9} & \textbf{35.0} & \textbf{33.8} & \textbf{54.7} & 82.2 & \textbf{14.0} & \textbf{66.9} & \textbf{74.4 }& \textbf{33.2} & \textbf{45.7} \\
& Remain \% & - & \textcolor{gray}{36\%} & \textcolor{gray}{34\%} & \textcolor{gray}{36\%} & \textcolor{gray}{38\%} & \textcolor{gray}{36\%} & \textcolor{gray}{40\%} & \textcolor{gray}{36\%} & \textcolor{gray}{38\%} & \textcolor{gray}{37\%} & \textcolor{gray}{40\%} & \textcolor{gray}{37\%} \\
\hline
\end{tabular}
}
\label{tab:beir-eval}

\end{table*}

\subsection{A Mean Error Perspective on First-$p$ Pruning}
\label{sub:mean-error-perspective}

\begin{figure}
    \centering
    \includegraphics[width=\linewidth]{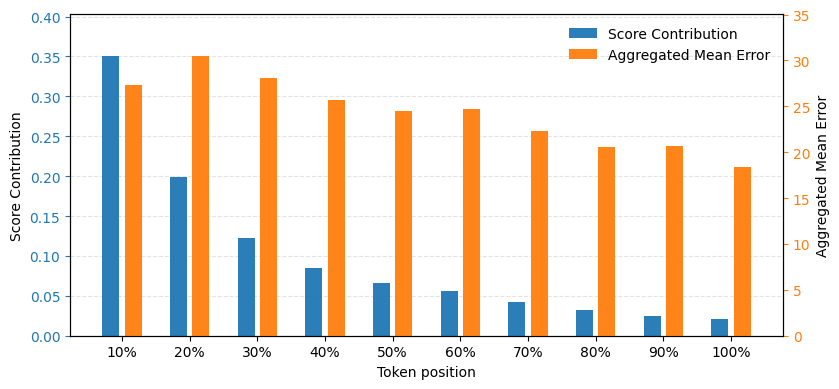}
    \caption{Distributions showing how often document tokens contribute to max-dot-product scores (in blue)
    and the aggregated mean errors over the position of tokens (in orange).}
    \label{fig:score-mer-k}
\end{figure}

The first-$p$ pruning strategy \citep{liuAnalysisMatchingMechanisms2024} is motivated by the observation that tokens appearing earlier in a sequence often contribute most to ColBERT scores for a fixed set of queries. While this method works reasonably well in practice, its underlying assumptions and limitations can be better understood through our Mean Error framework:
    \paragraph {Mean Error vs. Score Contributions.} First-$p$ pruning estimates token importance based on maximum dot-product contributions and observes that earlier tokens tend to be more influential. However, when we measure token importance using a sample of MS~MARCO's train set as described in \cite{liuAnalysisMatchingMechanisms2024}, the mean error calculated across sampled query vectors, a more relevant metric for analyzing pruning, shows that the distribution is much less skewed. While early tokens can have higher individual contributions, the mean error reveals a relatively more balanced view of which tokens can be safely pruned, as illustrated by Figure~\ref{fig:score-mer-k}.

    \paragraph{Non-iterative nature of First-$p$ pruning.} First-$p$ pruning applies a static selection of tokens without updating for how the contributions of remaining tokens change as others are removed. Following our discussion in Section~\ref{sec:method}, it is clear that First-$p$ pruning performs a single-step, non-iterative pruning. Calculating mean errors after pruning every token reveals why this non-iterative approach of First-$p$ pruning is suboptimal.
    
\begin{figure}
    \centering
    \includegraphics[width=\linewidth]{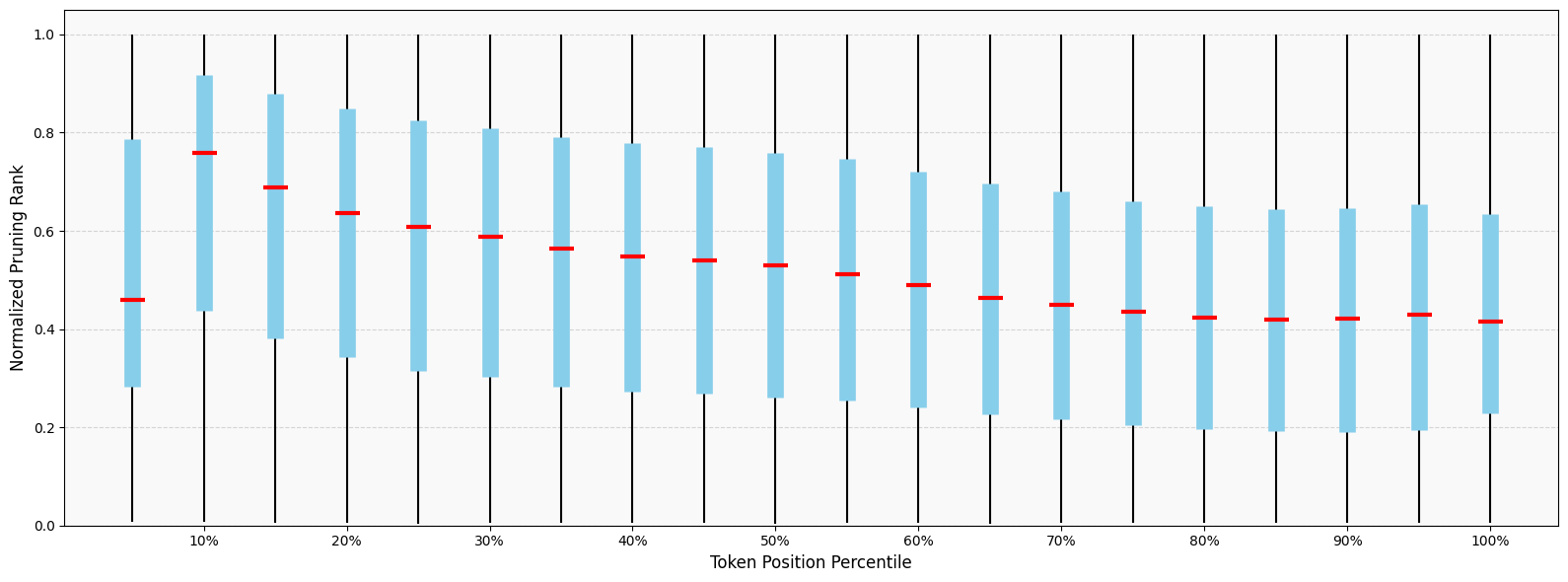}
    \caption{Distribution showing when a token at a particular position is pruned, relative to other tokens in the document. Lower values signify earlier pruning.}
    \label{fig:prune-candles}
\end{figure}

     \paragraph{Pruning order analysis.} Our framework allows a more detailed view of token importance by computing a pruning rank for each token within its document, normalizing ranks by document length, and grouping tokens by their relative position percentile. Aggregating ranks across the corpus and visualizing them using candlestick charts highlights the variability in token importance with respect to token position, revealing that earlier tokens do seem to be pruned later, implying that they are of moderately higher importance. Additionally, we also see that the first bin in particular is susceptible to early pruning, thus informing us that the first tokens in general are of lower importance (Figure~\ref{fig:prune-candles}).

\subsection{Guiding Pruning Decisions with Mean Error}
\begin{figure}[t]
    \centering
    \begin{subfigure}[t]{0.8\linewidth}
        \centering
        \includegraphics[width=\linewidth]{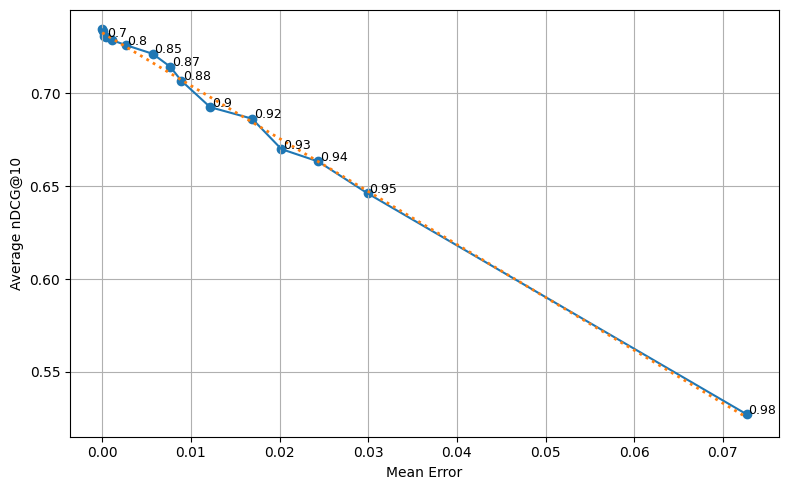}
        \caption{Average over TREC-DL19 and TREC-DL20}
        \label{fig:mer-perf}
    \end{subfigure}
    \hfill
    \begin{subfigure}[t]{0.8\linewidth}
        \centering
        \includegraphics[width=\linewidth]{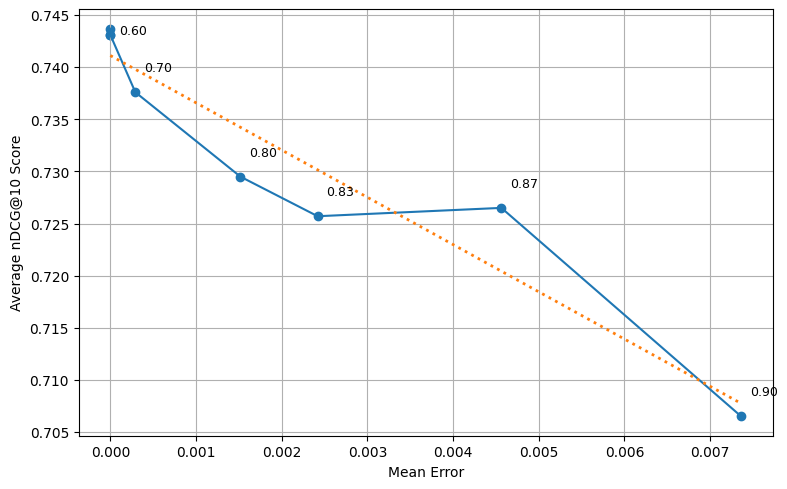}
        \caption{TREC-COVID}
        \label{fig:mer-perf-tc}
    \end{subfigure}
    \caption{Relationship between Mean Error and retrieval performance for \colbertvtwo{-sim} ($\alpha=0.8$) embeddings on different datasets. Orange dotted lines denote the line of best fit.}
    \label{fig:mer-perf-combined}
\end{figure}

We study the relationship between the \emph{Mean Error} induced in document representations and retrieval effectiveness under varying pruning ratios. Using Voronoi Pruning, we observe a strongly linear relationship between ME and nDCG@10, with an $R^2$ value of $0.99$ across TREC-DL19 and TREC-DL20, and $0.91$ on the TREC-COVID dataset (Figure~\ref{fig:mer-perf-combined}). This indicates that ME serves as a highly reliable proxy for post-pruning retrieval performance.

This strong correlation enables a practical pruning strategy: rather than tuning pruning ratios directly against retrieval metrics, one can select a target ME threshold to achieve a desired efficiency–effectiveness trade-off. Importantly, this behavior aligns with our geometric interpretation of ColBERT scoring-pruning that induces low ME preserves high-volume Voronoi regions that dominate similarity aggregation, resulting in minimal performance loss. As a result, Mean Error provides a principled and computationally efficient criterion for guiding pruning decisions.

\jlrannote{Conclusion??}
\section{Conclusion}
We introduced Voronoi Pruning, a principled framework for token pruning in late-interaction retrieval models grounded in the geometry of embedding spaces. By casting pruning as the minimization of expected retrieval error and interpreting token importance through Voronoi regions, our approach directly optimizes the objective that prior methods approximate indirectly.

Across in-domain and out-of-domain evaluations, Voronoi Pruning consistently achieves strong effectiveness–efficiency trade-offs, matching or outperforming learned pruning methods while remaining entirely post hoc and orders of magnitude faster than prior optimization-based approaches. Crucially, our method remains robust under aggressive pruning for in-domain tasks, where heuristic and dominance-based strategies degrade sharply.

Beyond its empirical performance, Voronoi Pruning provides a powerful framework for understanding token-level relevance in late-interaction models, enabling both principled pruning and new analytical insights into existing heuristics.

That said, our framework also has important limitations. Since our pruning criterion relies on mean error, which captures the average degradation in maximum dot-product similarity, it does not account for heterogeneous error distributions across the query space. In practice, small regions with large localized errors may be masked by low expected error, even if they are important for certain queries. Also, our objective optimizes for the preservation of maximum inner-product scores, which, while closely tied to retrieval performance, is only a proxy for downstream information retrieval loss. Finally, our approach is purely selection-based. It identifies the best subset of token representations within a fixed embedding space, but does not modify or optimize the space itself to make it inherently more prunable. 

Addressing these limitations opens several promising directions for future work. In particular, our geometric formulation could be extended to incorporate distribution-aware or worst-case error objectives, tighter alignment with task-level retrieval losses, or learning procedures that explicitly shape embedding spaces to increase their prunability. This could potentially aid in developing prunable models for out-of-domain data retrieval, where robustness under domain shifts remains a challenge. We leave these directions to future work and hope that the framework introduced here enables more principled analysis, probing, and optimization of embedding spaces in neural retrieval systems.

\begin{acks}
The authors acknowledge the peoples of the Woi Wurrung and Boon Wurrung language groups of the eastern Kulin Nation on whose unceded lands ACM SIGIR 2026 was hosted. We pay our respects to their Elders past and present, and extend that respect to all Aboriginal and Torres Strait Islander peoples today and their continuing connection to land, sea, sky, and community.

This work was supported by resources from the Institut du Développement et des Ressources en Informatique Scientifique (IDRIS) under the GENCI allocation AD011015851. The authors acknowledge the Agence nationale de la recherche (ANR – FRANCE) for its financial support of the GUIDANCE project n°ANR-23-IAS1-0003, the SEMIAMOR project ANR CE23-2023-0005, and the ANR Cluster IA ANR-23-IACL-0007. 
\end{acks}

\bibliographystyle{ACM-Reference-Format}
\balance
\bibliography{references}
\end{document}